# Gigahertz frame rate imaging of charge-injection dynamics in a molecular light source

Anna Rosławska[1,2,*], Pablo Merino[1,3,4], Christopher C. Leon[1], Abhishek Grewal[1], Markus Etzkorn[1,5], Klaus Kuhnke[1,*], Klaus Kern[1,6]

[1] Max-Planck-Institut für Festkörperforschung, D-70569, Stuttgart, Germany.

[2] Université de Strasbourg, CNRS, IPCMS, UMR 7504, F-67000 Strasbourg, France.

[3] Instituto de Ciencia de Materiales de Madrid, CSIC, E-28049, Madrid, Spain.

[4] Instituto de Física Fundamental, CSIC, E-28006, Madrid, Spain.

[5] Institut für Angewandte Physik, TU Braunschweig, D-38106 Braunschweig, Germany.

[6] Institut de Physique, École Polytechnique Fédérale de Lausanne, CH-1015 Lausanne, Switzerland.

* roslawska@ipcms.unistra.fr

* k.kuhnke@fkf.mpg.de

**Abstract**

Light sources on the scale of single molecules can be addressed and characterized on their proper sub-nanometer scale by scanning tunneling microscopy induced luminescence (STML). Such a source can be driven by defined short charge pulses while the luminescence is detected with sub-nanosecond resolution. We introduce an approach to concurrently image the molecular emitter, which is based on an individual defect, with its local environment along with its luminescence dynamics at a resolution of a billion frames per second. The observed dynamics can be assigned to the single electron capture occurring in the low-nanosecond regime. While the emitter's location on the surface remains fixed, the scanning of the tip modifies the energy landscape for charge injection to the defect. The principle of measurement is extendable to fundamental processes beyond charge transfer like exciton diffusion.

**Main text**

Energy conversion in both artificial and natural systems proceeds via a sequence of fundamental processes occurring at the quantum level of single photons and single electrons. The individual charges during redox reactions, biosynthesis and light emission from optoelectronic devices undergo a series of processes like tunneling, hopping, or recombination leading to measurable chemical, electronic or optical signals[1,2]. In particular, the tunneling of the electron through short molecular bridges plays a crucial role in electron transport within molecular wires[3] and between nucleic acids[1]. While these charge transfer processes can be of a few picoseconds or less, they are typically an order of magnitude slower[2,4] when the transfer distances are increased to a few nanometers and are thus accessible on the sub-nanosecond temporal and controllable at the sub-nanometer spatial scales.

Probing charged species with such spatial control can be achieved using scanning tunneling microscopy (STM). This approach is sensitive to the electronic density of states which enables electronic spectroscopy and imaging of charged single atoms[5], molecules[6] and defects[7–10], including elucidation of intramolecular details if combined with atomic force microscopy[11]. These studies, however, investigated static systems, in which a charge was either permanently residing in the system or replenished faster than the time resolution of the measurement. Requiring the measurement of small currents in the pA range, the temporal resolution of STM is typically limited to millisecond resolution. This limitation can be overcome by using advanced methodologies like all-electronic pump-probe spectroscopy[12] or coupling with ultrafast laser pulses[13–15]. In such experiments, the signal is detected by employing the STM tunnel current to read out the averaged response of the system that varies with the delay between the applied pulses. Employing STM-induced luminescence (STML)[16] in contrast provides a specific selectivity to processes that result in photon emission, in particular to electroluminescence of molecular emitters[16–25], including charged species[17,18]. Thanks to time-resolved single-photon detectors, the STML signal can be probed with sub-nanosecond temporal resolution[22,23,26], albeit limited to local point measurements. In our work, we map the electroluminescence in the time-domain and record optical nanometer-nanosecond snapshots of light emitted by single defects in thin organic films that light up within a few nanoseconds after pulsed electronic excitation. Because photon emission is intimately linked to electron injection, the electroluminescence delay can be used as a real-time and real-space monitor of the occurrence of an individual nanosecond electron transfer process. The short lifetime of the intermediate excitonic state (< 1 ns), which converts the charge injection into light emission preserves the time resolution of the measuring principle. The injection rate depends on the position of the STM tip, which remote-controls the electric field at the defect.

The experiment is schematically presented in Fig. 1a. We study time-resolved STML (TR-STML) from individual defects in thin $C_{60}$ films[23–25,27,28] grown on an Au(111) substrate using a cryogenic STM with optical access. The light emission from the defect is periodically induced by high-fidelity 100 ns long square pulses[12,29,30] with sharp edges (~ 1 ns rise time) and an amplitude $U_{pulse}$ added to the static bias voltage. The response of the system is probed by recording TR-STML intensity transients, P(t), with sub-ns time-resolution (see Methods for more details), which reveal an exponential rise and decay with respect to the applied square pulse both encoding the time ($\tau_e$) a single electron takes to be captured in the defect from the substrate after a hole has been injected from the tip [25]. This approach relies only on the optical signal and does not require peculiar electronic configuration to access charge injection rates[29,31].

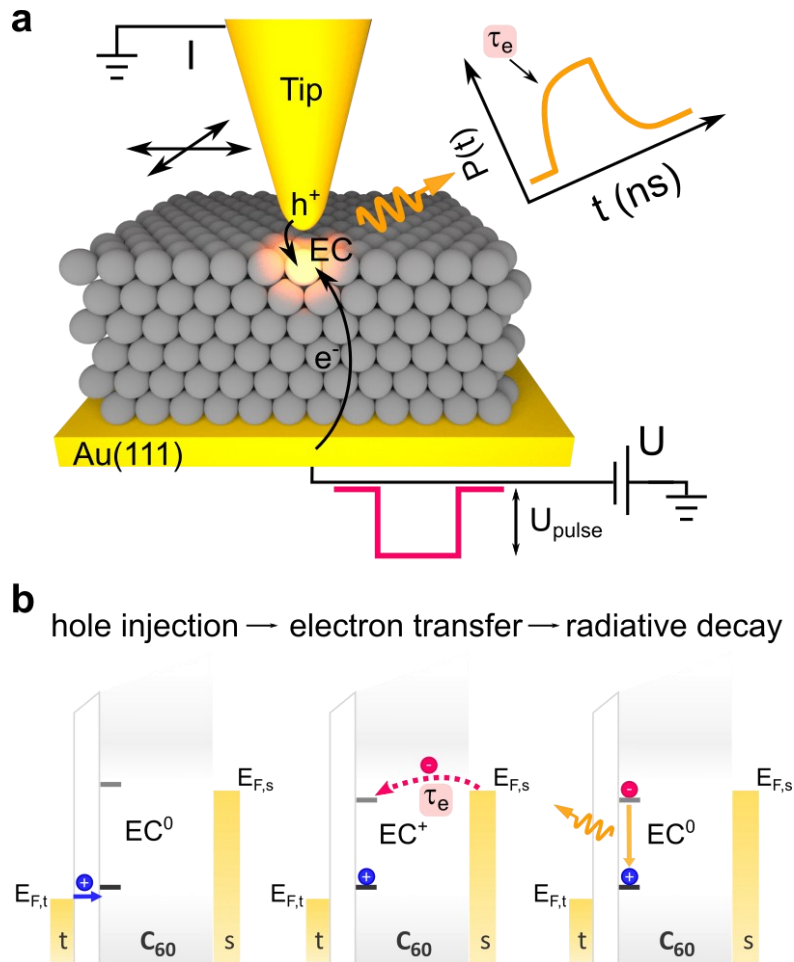

*Figure 1.* Probing the single electron injection dynamics. a) Experimental scheme. Negative nanosecond voltage pulses (100 ns long, 1 ns rise/fall time) are applied to an STM junction. The pulses enable photon emission from an individual emission center (EC) in a $C_{60}$ thin film, which is recorded as a function of the delay with respect to the pulse arrival. The measured exponential onset is a probe of the single electron injection time ($\tau_e$). By varying the tip position on the surface, we map $\tau_e$ with nm spatial precision. b) Energy diagrams illustrating the mechanism of the electroluminescence, which is a result of a sequence of events labelled on top of the panel. t = tip, s = Au substrate.

In a first step the geometry of the defect and its neighborhood are studied by STM topography. $C_{60}$ molecules are resolved internally revealing their individual orientation in the top layer of the thin (< 10 nm) film (Fig.2a). The simultaneously recorded electroluminescence yield at each pixel (photon map, Fig. 2b) shows the spatially-confined emission center (EC) localized around the molecule numbered as 1. It is known from earlier studies that ECs in $C_{60}$ are related to structural defects that trap a hole and an electron and enable emission, otherwise symmetry forbidden, from the lowest singlet electron-hole state (exciton) of $C_{60}$[23–25,27,28]. The optical emission spectra at increasing distance from the EC, measured on molecules with identical orientation, still show the characteristic emission lines of $C_{60}$ (Fig. 2c)[24]. The origin of this specific EC is further discussed in the Supplementary Information.

When the bias voltage of the STM is driven more negative by the transient voltage pulse than the applied static negative voltage it enables hole injection into a neutral defect state ($EC^0$) such that it becomes transiently charged ($EC^+$) (left panel in Fig.1b). This defect is then neutralized by a single electron transfer from the Au(111) substrate, which occurs within time $\tau_e$ (middle panel in Fig.1b). Note that the electron transfer from the substrate is substantially enhanced by the strong electrostatic potential of the trapped hole which shifts the electron defect level below the Fermi energy of the substrate ($E_{F,s}$)[25]. This process results in the creation of an electron-hole pair (exciton) at the defect that may decay radiatively by emitting a photon (right panel in Fig.1b). In this study, we apply voltage pulses of amplitude $U_{pulse}$ that move the Fermi level of the tip ($E_{f,t}$) from inside the bandgap between the states derived from highest occupied molecular orbital (HOMO) and the lowest unoccupied molecular orbital (LUMO) of the $C_{60}$ film to the highest lying HOMO states within 1 ns, switching on charge injection and subsequent photon emission.

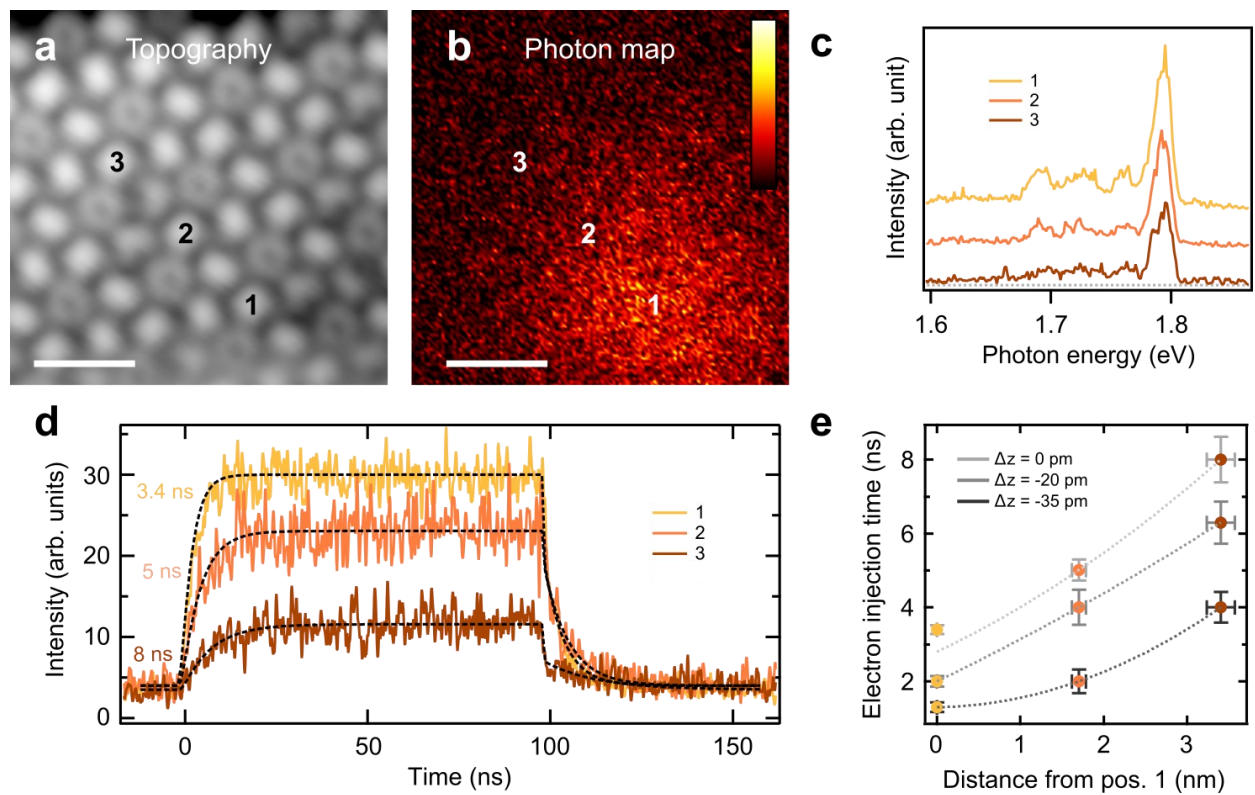

*Figure 2.* Electron transfer time to the defect as a function of the lateral tip position. a) Constant current STM topography U = -3 V, I = 30 pA, scale bar 2 nm. b) Electroluminescence yield map (photon map) recorded simultaneously with a). The color scale bar: 0 – 60 counts /s /pA. c) Optical spectra recorded at positions marked in a) and b). U = -3 V, I = 30 pA. The traces are offset for clarity; the dashed line indicates the baseline. d) TR-STML transients measured at positions marked in a) and b), average current during the pulse, $I_{pulse}$ = 12 pA, $U_{on}$ = -2.83 V, $U_{off}$ = -2.53 V. The dashed lines represent fits to the kinetic model, the extracted $\tau_e$ values are indicated next to the traces. e) Electron injection time measured at different horizontal positions (1-3) and different vertical offsets ($\Delta z$). $I_{pulse}$ = 12 pA ($\Delta z$ = 0pm), $I_{pulse}$ = 20 pA ($\Delta z$ = -20pm), $I_{pulse}$ = 37 pA ($\Delta z$ = -35 pm). The dashed lines are guides to the eye. The vertical error bars are the fitting errors, the horizontal error bar is 5% error of the distance measurement.

In the next step, we study in detail the dynamics of this electron injection process as a function of lateral position. We choose 3 molecules (numbered 1-3) with identical orientation (hexagon-hexagon bond facing upwards) where the same local density of states (LDOS) results in same tip height for a fixed tunneling current set point. This is particularly critical for comparing the respective dynamics at the molecules since the electron transfer rate depends strongly on the tip-sample distance[25]. For consistency, we show that the emission spectra are identical at all three positions (Fig. 2c) exhibiting no plasmonic contribution[28,32] and varying only in intensity. We record the electroluminescence transients at the marked positions and plot them in Fig. 2d. As observed directly from the TR-STML signal, the dynamics encoded in the rising and falling edges of the light pulse slow down when hole injection from the tip occurs farther from the center of the EC. Here, we would like to stress that the exciton recombination and photon emission are believed to always occur at the center of the EC, close to molecule 1. Only at this position are the selection rules relaxed and permit emission[24,27]. The exciton lifetime remains shorter than 1 ns and the hole injection is comparatively slow (µs regime) due to both the tunneling current and trapping efficiency being low[23,25], such that the majority of the current passing through the system will not contribute to the luminescence. Because these processes occur at time scales that are different from the one observed in the experiment, the dynamics observed in the luminescent transients can be related to the electron injection from the substrate. Its rate can be obtained by fitting the transient to a kinetic model describing the sequence shown in Fig. 1b[25,26]. In Fig. 2e we plot the extracted $\tau_e$ as a function of the distance from the central molecule (1) and find that it increases from 3 ns to 8 ns when moving away laterally by 4 nm. Additionally, in line with our previous observations, we find that $\tau_e$ decreases when the tip-sample distance is reduced (three curves in Fig. 2e)[25]. This can be ascribed to a reduction of the energy barrier at the $C_{60}$/Au(111) interface due to the increase of the electrical field.

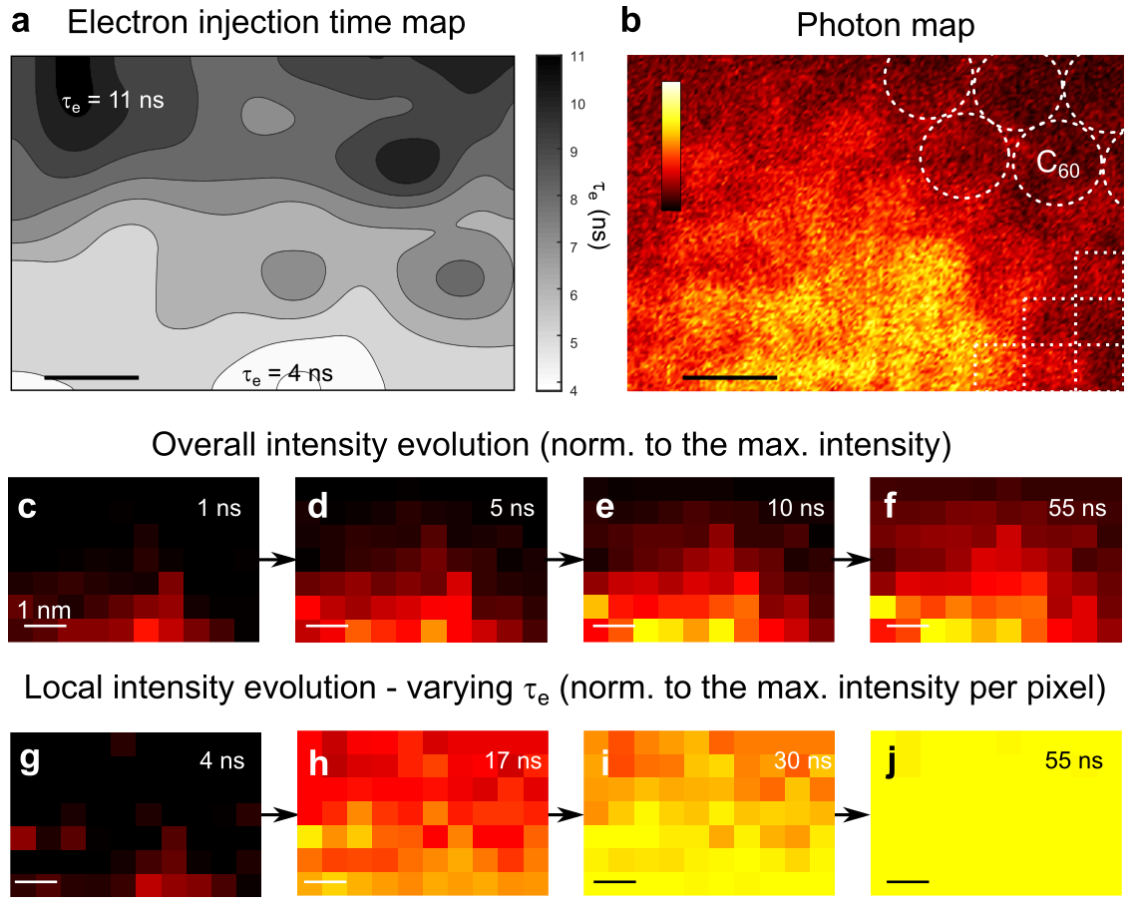

*Figure 3.* Electron injection dynamics as a function of the position on the surface. a) Map of electron injection. The time constants are extracted from TR-STML transients measured on a 10x7 grid (400 s integration time per pixel, $U_{on}$ = - 2.83 V, $U_{off}$ = -2.53 V) and spatially interpolated. b) Photon map of the same region as in a). An overlay of the $C_{60}$ lattice at the interface (grid) is represented by dashed circles (lines). The color scale intensity ranges from 0 to 10 kcts s$^{-1}$. c)-f) Light intensity snapshots extracted at the indicated time delays after the arrival of the pulse to the junction. The images are normalized to the maximal light intensity over the whole data set, which is presented in Supplementary Video 1. g)-j) Light intensity snapshots normalized to the maximum on each pixel reached after ca. 50 ns. The luminescence reaches its maximum faster at the center of the EC (lower part of the images). All scale bars are 1 nm.

Next, we extend our analysis by mapping the charge injection times as a function of the tip position near an EC by measuring TR-STML transients on a 10x7 points grid (see Methods). In Fig. 3a we plot a spatially interpolated map of extracted $\tau_e$ and in Fig. 3b we present the photon map to present the spatial extension of the EC for comparison. As observed in Fig. 2d, the electron injection time to $EC^+$ increases when the tip is located at the periphery of the EC.

The measurements described above can also be represented in a sequence of images showing the time evolution of light emission by slicing the 3-dimensional data block along constant delay times. At first, we compare the map of fitted steady-state intensity (Fig. 3f) that is usually reached after 30-40 ns (see Fig. 2d) with the photon map (Fig. 3b) of the same area. Indeed, the snapshot reproduces correctly the spatial extent of the EC, measured by the spectrally integrated photon map. Next, we present the snapshots for various delays in Fig. 3c-f, which are normalized to the highest recorded intensity within the whole data set. As expected from the lateral dependence of the charging time (Fig. 2), the light intensity evolves slower when the tip is positioned at the periphery of the EC and is direct visualization of the increase in the electron capture time. This time evolution is emphasized in panels g-j of Fig. 3, where each point has been scaled to its intensity maximum reached under steady-state conditions so that each pixel eventually reaches a value of 1. For instance, Fig. 3h,i demonstrate that in the central part of the EC (lower part of the images), the relative intensity is higher than at the peripheries (upper part of the images). By measuring the nanosecond-resolved light emission using TR-STML we can thus probe and follow the single charge transfer time in 4 dimensions at the molecular scale with GHz frame rates (intervals of 0.7 ns) as visualized in Supplementary Video 1.

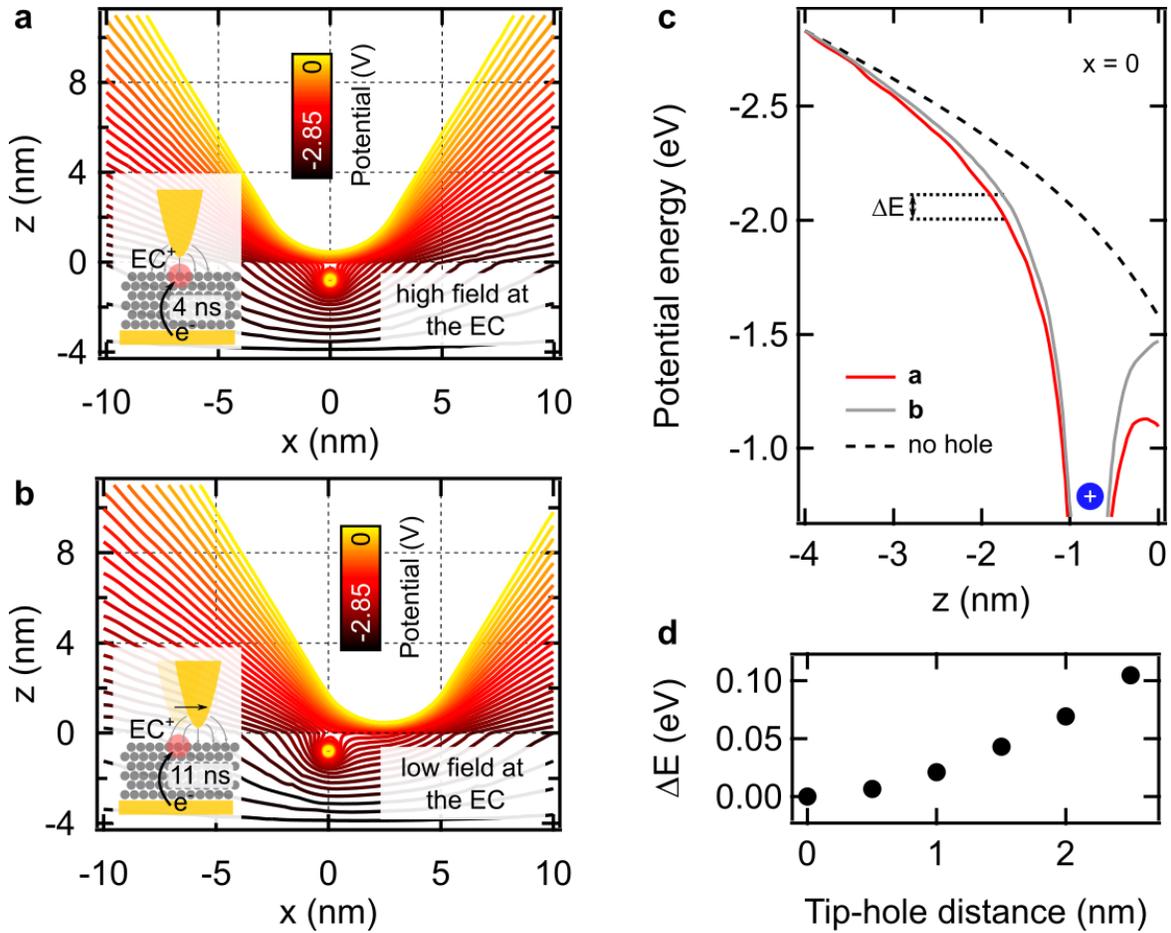

*Figure 4.* *The electric potential at EC⁺ as a function of the tip position. Two limit cases are presented, the tip placed at the center of the EC (a) and 2.5 nm away (b). (c) The potential cross-sections along the position x = 0. The dashed line plot shows the potential in the absence of the hole. (d) The increase of the potential barrier when the tip moves away from EC⁺.*

The observed reduction in the electron injection rate (i.e. increase in the charge transfer time constant) as a function of the distance from the central position of the EC can be explained by the electric field inside the $C_{60}$ film. The electron and hole trap states have the LDOS maximum in the center of the EC with the LDOS decaying with the lateral distance[27]. Thus, when the tip is localized at the periphery of the EC, the hole can still reach the defect state and create a charged defect state EC⁺. Additionally, the probability of trapping the hole at the defect is reduced compared to the central positions of the EC, because it is more likely that the hole will be transported through the semiconducting $C_{60}$ layer directly to the substrate. This results in a lower electroluminescence yield, as shown in Fig. 2d (number of counts per time bin) and Fig. 3b. The electron capture by $EC^+$ is purely field-driven, does not involve transport from the tip and thus constitutes a parameter independent of the emission intensity. It is induced at some distance away from the tip apex, similarly as the sharp rings observed in dI/dV maps which indicate local charging effects by the tip stray field[7–10]. The electric field is controlled by the tip position as confirmed by electrostatic calculations (see Methods) shown in Fig. 4. Remarkably, when the tip is located at the periphery of the EC, the electric field is reduced but still sufficiently strong at the $EC^+$ position, allowing the electron tunneling. In Fig. 4c we compare the potential energy situation for the cases in Fig. 4a and b and find that the potential energy barrier for the electron injection from the substrate is increased by 0.1 eV ($\Delta E$) when the tip is displaced 2.5 nm from the location of the hole. $\Delta E$ increases gradually when the tip is moved away from $EC^+$ as plotted in Fig. 4d, which slows down the exciton formation process (Fig. 2e). A similar energy barrier increase is observed when the tip is retracted from the surface above the defect[25], as shown in Fig. 2e.

Alternatively, one might assume that the exciton formation occurs always below the tip apex followed by exciton diffusion towards the defect where the exciton would radiatively decay. However, the diffusion process of the exciton occurs within its lifetime that is shorter than the resolution of the experiment. Even though this scenario is consistent with the intensity fall-off with distance from the defect it is not compatible with the observed slowed-down dynamics at a large distance. Exciton formation away from the defect would rather require a constant or even shorter electron injection time since in this scenario the hole would not get trapped and rapidly move to the substrate – driven by the strong electric field. Finally, we can exclude that the increased nanosecond time delay results from the diffusion of the hole injected by the tip at some nanometer distance from the defect, as the charge hopping time in $C_{60}$ was reported to be in the femtosecond regime[33].

In conclusion, we establish an approach to image a light emitter on the nanoscale and map the evolution of its light emission with a rate of $10^9$ frames per second. The observed luminescence evolution reflects the electron transfer from the substrate to a localized emitter and is controlled by the electric stray field of the STM tip, mimicking the energy landscape modifications induced by localized charges and different molecular species. While the overall charge transport in our system is dominated by the current passing through the HOMO-derived states, monitoring electroluminescence allows us to be sensitive to the electron injection to the defect only. We envision our approach to be used in future studies to explore single-electron injection dynamics that reach even sub-molecular resolution with single-molecule emitters or atomic point defects. In a material whose exciton lifetime is increased, for instance, based on triplet emission, the method presented here could be adapted to study exciton diffusion in real time.

**Methods**

Scanning tunneling microscopy-induced luminescence

All experiments were performed using a home-built ultra-high-vacuum low-temperature (4 K) STM with optical access provided by three lenses located in the STM head with their focus on the tip apex. We couple one of the resulting three independent light paths to a single-photon avalanche photodiode (SPADs, MPD-PDM-R) and another one to an optical spectrometer (spectrograph: Acton SP 300i, CCD camera: PI-MAX). The Au(111) crystal is prepared by repeated cycles of $Ar^+$ sputtering and annealing (up to 850 K). $C_{60}$ is thermally evaporated from a Knudsen cell (850 K) for 1h on the crystal held at room temperature.

Time-resolved measurements

Transmission function-corrected[25,30] voltage pulses (2 MHz repetition rate, 100 ns length, 1 ns rise/fall time) are produced by an arbitrary wave generator (AWG, Agilent M8190A) and sent to the tunnel junction through high-frequency optimized wiring (semi-rigid and coaxial cables). The amplitude of the pulses is -300 mV, which is added to the DC offset bias ($U_{off}$) by a bias tee (Picosecond Pulse Laboratories, 5550B). $I_{pulse}$ is defined as the tunneling current measured at $U_{on}$ for the same tip-sample distance as during the pulse measurement. For the measurements shown in Fig. 3, the feedback loop was off during acquisition of the transient but turned on between the measurements to correct for the z drift of the STM tip. To minimize the overall drift, the tip was stabilized at the EC for 10 h before the series. The integration time per point was 400 s. More details on the measurement can be found in the Supplementary Information.

Electrostatic calculations

The calculations are done with the Mecway finite element analysis software (Mecway Ltd., New Zealand) in the full 3D geometry of the problem. The results are represented in the figures by a cut along the symmetry plane of the geometry defined by the tip axis and the position of the charge. For details see Supplementary Information.

**Acknowledgements**

We would like to thank O. Gunnarsson and G. Schull for fruitful discussions. A.R. acknowledges support from the European Research Council (ERC) under the European Union's Horizon 2020 research and innovation program (grant agreement No 771850) and the European Union's Horizon 2020 research and innovation programme under the Marie Sklodowska-Curie grant agreement No 894434. P.M. acknowledges support from the A.v. Humboldt Foundation, the ERC Synergy Program (grant no. ERC-2013-SYG-610256, Nanocosmos), Spanish MINECO(MAT2017-85089-C2-1-R) and the "Comunidad de Madrid" for its support to the FotoArt-CM Project S2018/NMT-4367through the Program of R&D activities between research groups in Technologies 2013, cofinanced by European Structural Funds.

**Author contributions**

A.R., P.M., C.C.L. and A.G performed the experiments. A.R. analyzed the data. K. Ku. performed the electrostatic calculations. M.E., K.Ku., and K.Ke. conceived and supervised the project. All authors contributed to discussions and writing of the manuscript.

**Supplementary information for:**

**Gigahertz frame rate imaging of charge-injection dynamics in a molecular light source**

Anna Rosławska[1,2,*], Pablo Merino[1,3,4], Christopher C. Leon[1], Abhishek Grewal[1], Markus Etzkorn[1,5], Klaus Kuhnke[1,*], Klaus Kern[1,6]

[1] Max-Planck-Institut für Festkörperforschung, D-70569, Stuttgart, Germany.

[2] Université de Strasbourg, CNRS, IPCMS, UMR 7504, F-67000 Strasbourg, France.

[3] Instituto de Ciencia de Materiales de Madrid, CSIC, E-28049, Madrid, Spain.

[4] Instituto de Física Fundamental, CSIC, E-28006, Madrid, Spain.

[5] Institut für Angewandte Physik, TU Braunschweig, D-38106 Braunschweig, Germany.

[6] Institut de Physique, École Polytechnique Fédérale de Lausanne, CH-1015 Lausanne, Switzerland.

* roslawska@ipcms.unistra.fr

* k.kuhnke@fkf.mpg.de

**Size of the emission centers**

For defects located in the subsurface layers, the apparent size (bright area on the surface) of the emission center (EC) can be even larger than the size of the ECs located at the top layer. Already for such an EC, even though the local density of states (LDOS) of the hole trap state is the highest in the central area of the EC, there is still experimentally measurable increased LDOS signal at a lateral distance of 2-3 nm from the center[1]. In the case of a subsurface defect, a hole transported through the film may hop sideways due to the hexagonal stacking, resulting in a larger efficient lateral distance from which the hole can be injected into the defect. Fig. S1 shows such variation in sizes of the ECs, which can have a diameter ranging from 1-2 molecules to 5-6.

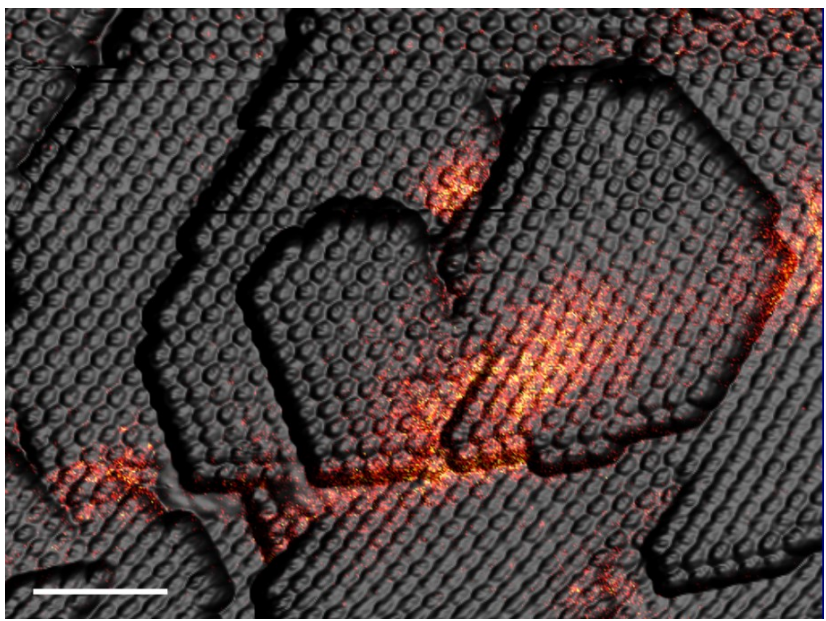

*Figure S1. Topography overlaid with electroluminescence yield map. U = - 3 V, I = 30 pA. Scale bar 5 nm. Intensity range (0-3 kcts s$^{-1}$).*

**Details of the characterization of the single electron injection dynamics in Fig. 2**

Origin of the EC

The origin of the EC presented in Fig. 2 of the main text may be related to a defect located on the surface or in one of the top layers of the $C_{60}$ film. Some of the ECs are related to a molecular misorientation (rotation) of molecules in the low-temperature $C_{60}$ (2x2) orientational superstructure[1], which is indicated in Fig S2a (the same area as in Fig. 2a in the main manuscript). In the case presented in Fig. 2, there are some perturbations in the superstructure in positions slightly off-center of the EC as marked in Fig. S2a by red dots. They are, however, not located at the central position where the defect trapping the charges and excitons is expected to be[1]. Thus, it is likely that the defect leading to split-off states is located below the surface. It is further supported by the fact that there are no split-off states visible in the scanning tunneling spectroscopy at position 1 (Fig. S2b) and only the onsets of the $C_{60}$ bands are resolved.

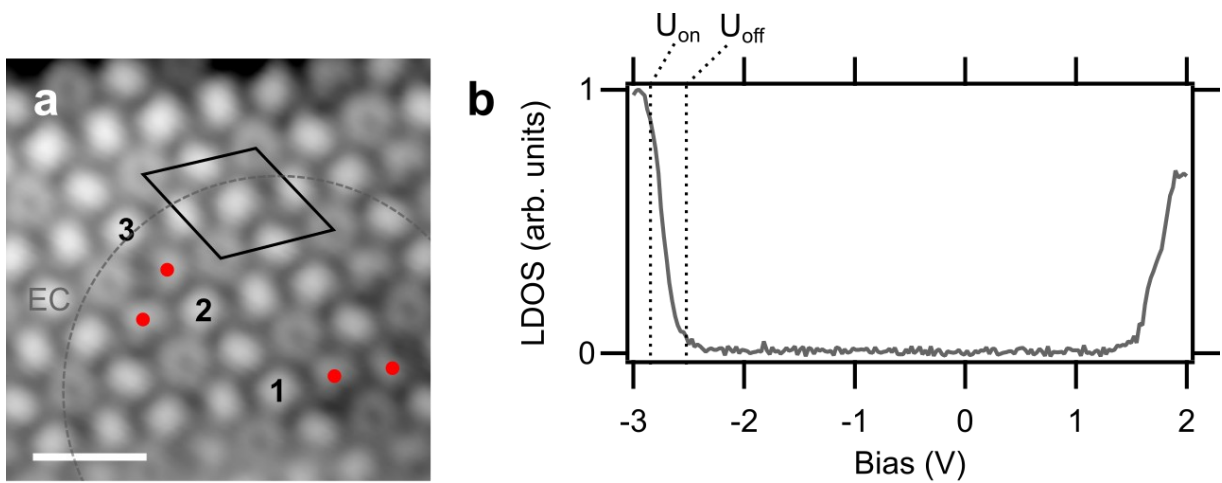

*Figure S2. Detailed characterization of the EC. a) STM constant current topographic scan of the same area as in Fig. 2a of the main text, U = -3 V, I = 30 pA. The spatial extent of the EC is marked by a dashed circle, the (2x2) superstructure is marked by a black rhombus, red circles mark the molecules that have a perturbed orientation. Scale bar 2 nm. b) Scanning tunneling dI/dV spectrum measuring the LDOS at position 1 in Fig. S2a, set-point: U =- 3 V, I = 30 pA.*

**Details of the grid measurement (Fig. 3)**

The measurements presented in Fig. 3 have been performed on a 10x7 rectangular grid corresponding to an area of 5x3.7 nm$^2$ with an integration time 400 s per pixel, 8 h for the whole data set. The three last points of the grid (upper right corner) are discarded due to a tip change (likely picking up a C$_{60}$ molecule from the layer) and replaced by the value of the last point before the change occurred. This operation preserves the spatial characteristics of the dynamics as evidenced in Fig. 3a of the main text.

Fig. S3 shows two traces demonstrating two representative cases from the dataset of the registered dynamics together with the fits to the kinetic model yielding electron injection time constants of 4.8 ns and 12 ns respectively.

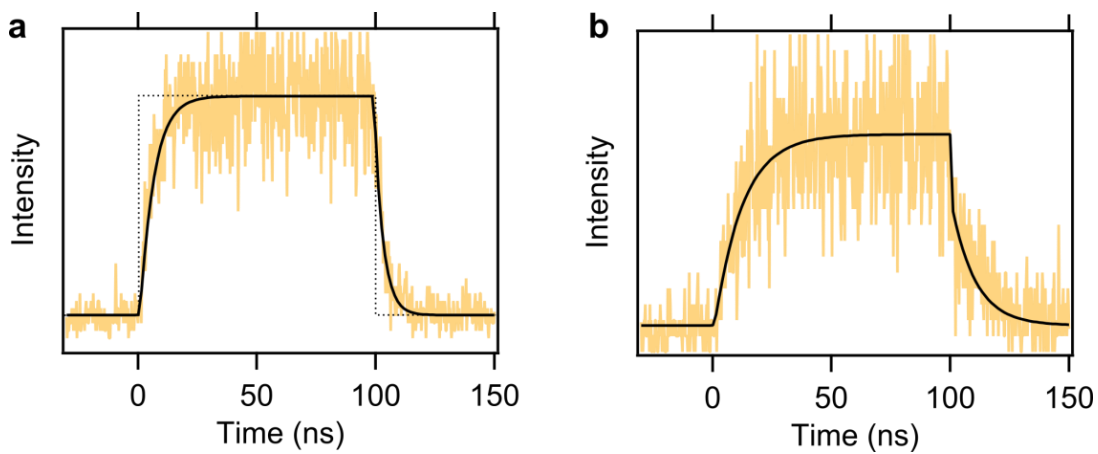

*Figure S3. Two representative transients obtained within the grid measurement. The yellow trace is the raw data; the black curve is the fit to the kinetic model. a) $\tau_{el}$ = 4.8 ns b) $\tau_{el}$ = 12 ns. The dotted line in a) indicates the shape and sharpness of the applied voltage pulse.*

The electron injection dynamics is present also in the falling edge (100-120 ns) of the transient. In that situation, the hole captured in the defect when the voltage pulse was present in the junction remains after the voltage returns to a value inside the bandgap. The hole shifts the electron trap sufficiently below the Fermi level of the substrate such that the electron can be still injected, and its dynamics mapped. Such a map is presented in Fig. S4b and shows a similar spatial dependence in which the measured time constant is slower in the peripheries of the EC. Due to the lower electric field (lower absolute bias voltage when the pulse is over), the time constants are relatively longer as compared to the rising edge (Fig. S4a). However, in that case, the electron injection dynamics is convoluted with another transport process in which the hole may tunnel to the tip (hole detrapping). It occurs on a similar time scale[2] and reduces the dynamical contrast. Therefore, the time constant map extracted from the rising edge (0-20 ns) of the transients (Fig. 3a) provides a more direct measure of the single electron injection dynamics. For a clear comparison, we show the maps in a raw, non-interpolated form (Fig. S4).

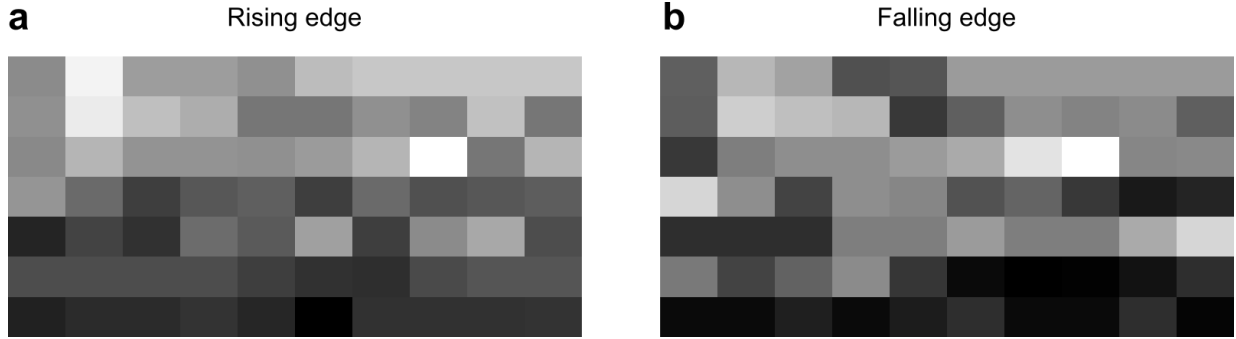

*Figure S4. Comparison between the electron injection time constant extracted from the rising (a) and falling (b) edges of the pulse. The studied area is the same as in Fig. 3 of the main text. Time constant ranges (black-white color scale): a) 3 ns – 13 ns. b) 7.1 ns – 25 ns.*

To correct for the z drift, we enable the feedback loop in-between measurements, which may result in a slight variation of tip-sample distance and affect the measured dynamics, similar to the data presented in Fig. 2e of the main text. To verify that our approach is correct we plot the relative z displacement at every position and present it in Fig. S5. Comparing Fig. S5 with Fig. S4 yields no correlation, so the variations in the tip-sample distance do not affect the general trend of the dynamics (in periphery of the EC the dynamics are slower). We note that these variations as well as to local changes in the topography, electronic structure, the non-symmetric tip shape may be responsible for the local non-monotonicity of the trend as visible in Fig. 3a and Fig S5a.

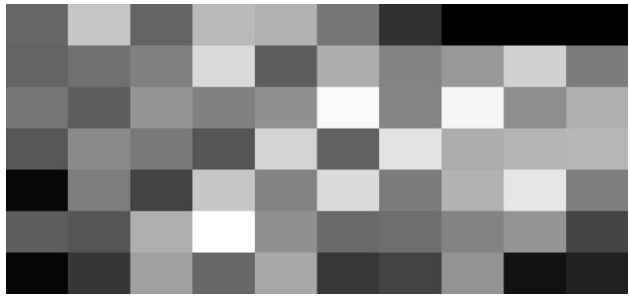

*Figure S5. Z displacement during the measurement. The studied area is the same as in Fig. 3 of the main text. Relative displacement range (black-white color scale): 0 – 115 pm.*

Finally, in Supplementary Video 1, we show a continuous evolution of the snapshots presented in Fig. 3: an "ns-nm video" covering the rising edge of the transients with subsequent snapshots every 0.7 ns. The video illustrates the evolution of light emission as a function of the tip position and is a real-time measure of the electron injection dynamics.

**Single-electron injection dynamics**

Fig. 2e of the main manuscript presents the time constants of the electron injection ($\tau_e$) extracted from the rising edge (approx. first 20 ns) of the transients. The fits to the kinetic model also yield the $\tau_e$ embedded in the falling edge of the transient, which is plotted in Fig. S6. Similarly to the tendency observed in Fig. 2e, the electron injection is slower when the tip is located at the peripheries of the EC. The time constants are longer for the falling edge because the bias (thus, the electric field) is lower and the electron injection is less favorable. We note that for the measurements presented in this work the voltage between the pulses ($U_{off}$) is already close to the onset of the highest occupied molecular orbital (HOMO) band, meaning that hole injection and light emission may take place between the pulses[2]. However, the emission intensity for t < 0 ns and t > 150 ns is very low and does not affect the exponential fit and the obtained time constant.

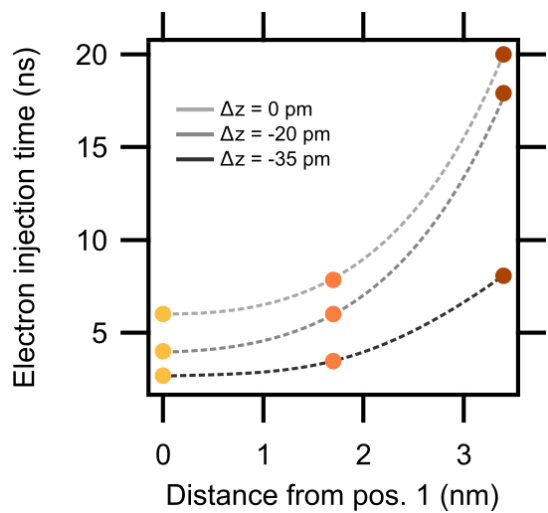

*Figure S6. Electron injection time extracted from the falling edges of TR-STML transients recorded at positions 1-3 marked in Fig. S2a. $I_{pulse}$ = 12 pA ($\Delta z$ = 0 pm), $I_{pulse}$ = 20 pA ($\Delta z$ = -20 pm), $I_{pulse}$ = 37 pA ($\Delta z$ = -35 pm), $U_{on}$ = -2.83 V, $U_{off}$ = -2.53 V. The dashed lines are guides to the eye.*

**Electrostatic potential calculations for a charge in the $C_{60}$ layer below the STM tip**

Electrostatic calculations were performed on a discrete grid using the Mecway finite element analysis software [Mecway Ltd., New Zealand] in the full 3D geometry of the problem making use of the mirror plane defined by the tip axis and the charge position which allows reducing the calculation to one half space. The grid consists of a total of 34367 elements with unevenly spaced points with the point density substantially increasing towards the tunnel junction and the charge position (see Fig. S7). The tip electrode consists of 1440 points set to 0 V potential with a conical shape (opening angle 60°) merged with a spherical shape of radius 3 nm. The distance from the tip apex to the plane $C_{60}$ surface is 0.5 nm, the thickness of the $C_{60}$ film (dielectric constant $\varepsilon=4.4$) is 4.0 nm. The elementary charge ($Q = 1.6\text{e-}19$ C) is placed at 0.8 nm depth below the surface inside the $C_{60}$ film. The plane substrate electrode below the $C_{60}$ film is defined by 688 points set to a potential of -2.85 V. For the presentation of the results potential values on the point charge and on its directly neighboring points have been removed as they tend to exhibit artefacts.

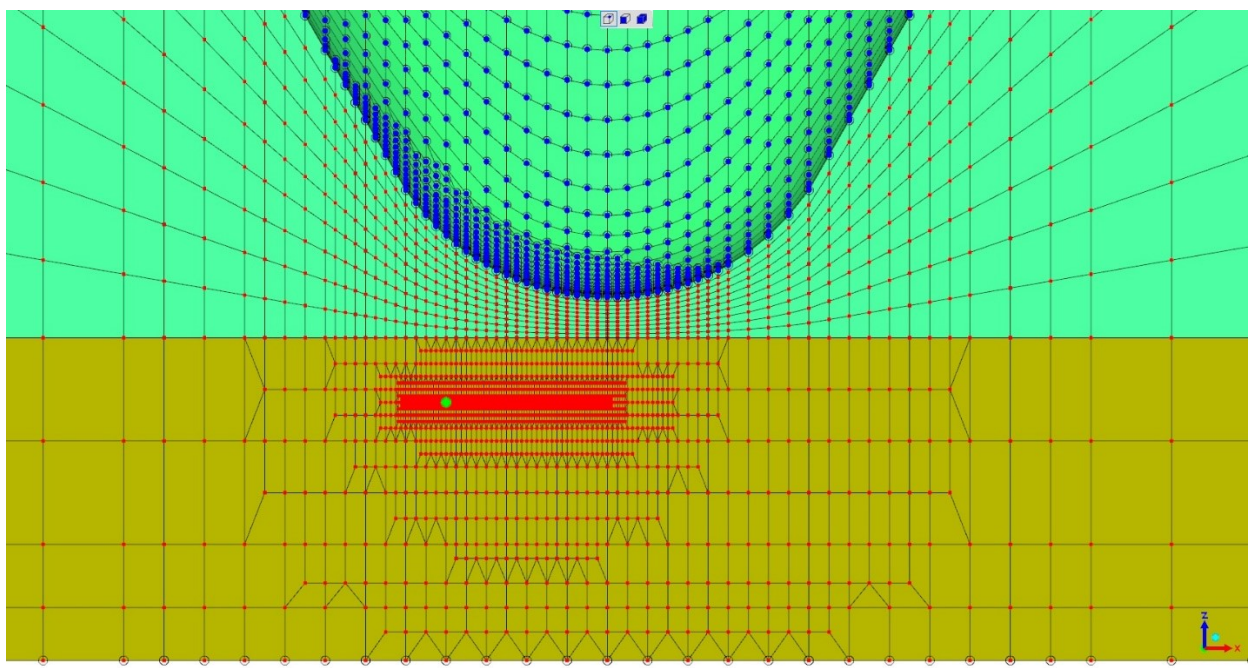

*Figure S7. Mecway plot of the 2-dimensional projection of the discrete grid for the calculation of the electrostatic potential in the STM junction with an elementary point charge (green dot) buried within the $C_{60}$ film. The projection reveals the complete grid of the (hollow) tip electrode.*

**Supplementary references**